\documentclass[runningheads]{llncs}
\usepackage[T1]{fontenc}
\usepackage{graphicx}
\usepackage{comment}
\usepackage{xcolor}
\usepackage{fvextra}
\usepackage{tikz}
\usepackage{amsmath}
\usepackage{amssymb}
\usepackage{booktabs}
\usepackage{longtable}
\usepackage{rotating}
\usepackage{tabularx}
\usepackage{array}
\usepackage{ragged2e}
\usetikzlibrary{positioning,fit,arrows.meta,calc}

\begin{document}

\title{RASER: Resilient Agent Scheduling and Execution Runtime for HPC Clusters}

\titlerunning{RASER: Resilient Agent Scheduling and Execution Runtime}

\author{Sima Attar-Khorasani\inst{1,3}\orcidID{0000-0002-7396-1983} \and
Matthias Lieber\inst{1,3}\orcidID{0000-0003-3137-0648} \and
Siavash Ghiasvand\inst{2,3}\orcidID{0000-0001-6627-0159}}

\institute{Center for Information Services and High Performance Computing (ZIH)\and
Center for Scalable Data Analytics and Artificial Intelligence (ScaDS.AI)\and
TUD Dresden University of Technology, Dresden, Germany \\
\email{first.last@tu-dresden.de}}

\maketitle              

\begin{abstract}
The emergence of modern agents powered by large language models has created a demand for executing long-horizon, autonomous workflows in various domains that require significant computational resources.
While High Performance Computing clusters provide the ideal infrastructure for these computation-intensive workloads, traditional HPC job schedulers such as Slurm are not designed for dynamic, agentic workflows characterized by unpredictable task durations, external API calls, and fault tolerance requirements of modern agents.
This work presents RASER, a user-space framework that enables seamless execution of agentic workflows on production HPC clusters by extending Slurm's internal primitives.
RASER introduces agentic job arrays with work stealing via shared filesystem queues, user-space checkpointing through application-level state serialization combined with Slurm requeue, and Apptainer container-based isolation without requiring any image modifications.
Evaluations demonstrate that RASER reduces makespan by nearly 39\% compared to static partitioning while achieving near-full CPU utilization.
RASER provides resilience against preemption and failures while maintaining minimal checkpoint/restore overhead.
It requires no kernel privileges or external database infrastructure, making it an accessible solution for deploying agentic workflows on existing HPC infrastructure.

\end{abstract}

\section{Introduction}
Current agents, powered by Large Language Models (LLMs), can execute long-horizon reasoning workflows that can span from minutes to hours.
Due to their abilities to autonomously divide complex objectives into manageable tasks, integrate with external tools and APIs, and iteratively improve their outputs, the use of agentic workflows shows an increasing demand in various fields~\cite{johnston26}.
Those decomposed tasks in agentic workloads however, in many cases still remain computation-intensive and require high-end computational hardware, similar to High Performance Computing (HPC) clusters.
Agentic workloads align naturally with the HPC clusters batch-oriented job model.
Within research environments, HPC clusters are a primary computation infrastructure;
utilizing these existing infrastructure for the modern agentic workflows eliminates duplicate deployment and maintenance efforts.
Using exclusive GPU and network resources that HPC clusters offer without any containerization overhead also avoids performance unpredictability, and enables latency-sensitive distributed operations.
Therefore, executing agentic workflows on HPC clusters typically represents the optimal approach.
It is important to mention that cloud-native approaches such as Kubernetes\cite{burns2016borg} stand out where agentic workflows become continuous services;
in such scenarios the overhead of managing a Kubernetes cluster deployed on the existing HPC infrastructure can be justified for productized agent platforms.

However, executing agentic workflows on HPC clusters presents unique challenges.
Unlike traditional HPC jobs, agent's workloads are often long-running and potentially fault-prone.
Agents often perform unpredictable tool invocations and external API calls that contradicts the static resource provisioning and restrictive network policies typical for HPC environments.
On HPC clusters, preemption is a common practice for shared resources, and heterogeneous tasks (e.g., 5-minute vs. 2-hour tasks) may create severe load imbalance.
Security and multi-tenancy concerns also increase when agent ensembles operate autonomously, accessing sensitive data or executing untrusted code, requires sandboxing solutions that preserve bare-metal performance without introducing the same overhead that motivated HPC deployment in the first place.
Currently, Slurm~\cite{yoo2003slurm} powers most of the HPC clusters in research environment~\cite{riddle2023slurm}.
It provides job arrays, that support static partitioning, and other primitives including \texttt{requeue} for long-running simulations and containers integration for dependency isolation.

This work introduces RASER, a framework that enables seamless execution of agentic workflows on production HPC clusters without any modification to the underlying software stack.
RASER combines and extends Slurm's primitives into a cohesive architecture that treats the Slurm batch scheduler as a dynamic agent substrate.
It provides agentic job arrays, enabling work stealing across array tasks via shared filesystem queues.
To the best of our knowledge this is the first usage of native Slurm job arrays for dynamic agent ensembles (non-parametric workloads).
Furthermore, the user-space checkpointing is achieved via application-level serialization of agent reasoning state combined with Slurm requeue.
Apptainer~\cite{kurtzer2017singularity} containers with bind-mounted runtimes provide per-agent isolation without requiring container image modification.
RASER runs in user-space and requires no kernel privileges or external database infrastructure.
Evaluating RASER against static partitioning demonstrates approximately 40\% faster return times while achieving near-full CPU utilization.
The RASER source code including sample plugins is available on Gitlab\footnote{\url{https://gitlab.hrz.tu-chemnitz.de/siat527e--tu-dresden.de/raser}}.
It is important to emphasise that RASER's primary objective is to democratize performant execution of agentic workloads on Slurm-powered HPC clusters; making this capability accessible to all users despite the limitations and barriers typical of production environments.

\section{Related Work}
\label{sec:related_work}

Efficient execution of heterogeneous agent ensembles on traditional HPC clusters is a challenging task due to multiple criteria including the nondeterministic behavior of agents.
A robust and efficient execution requires dynamic scheduling, fault tolerance, checkpointing, and secure isolation.
In most cases, altering the software stack on production HPC clusters is not feasible and high privilege execution is also not permitted.
Although the execution of complex computational workflows on High Performance Computing (HPC) clusters has been extensively studied, with various systems proposing different approaches to workflow management, resource scheduling, fault tolerance, and containerization, none of them provide a complete solution that satisfies all these criteria.
Therefore, this section mainly focuses on previous works that addressed relevant parts required for realizing such framework.

Several systems target heterogeneous HPC workloads.
RADICAL-Pilot integrates PMIx/PRRTE for process management although requires external database infrastructure and complex middleware~\cite{merzky2022pilotpmix}.
Parsl~\cite{babuji2019parsl} provides Python-based parallelism scaling to thousands of nodes, yet relies on static provisioning without native work-stealing.
Dask~\cite{rocklin2015dask} offers dynamic task scheduling however, is designed for data analytics rather than long-running agentic workflows.
Balsam~\cite{salim2019balsam} automates dynamic workflow scheduling with a pilot-based launcher but lacks integration with Slurm job arrays and application-level checkpointing.
TaskVine~\cite{sly-delgado2023taskvine} excels at data-intensive workflows but does not address LLM agent requirements such as stateful reasoning and unpredictable tool invocations.

For optimized utilization of HPC hardware, Ray~\cite{moritz2018ray} provides distributed primitives for AI applications.
Singularity/Apptainer~\cite{kurtzer2017singularity}, as a widely used container solution, enables portable HPC containers without root privileges.
Shifter~\cite{gerhardt2017shifter} and Charliecloud~\cite{priedhorsky2017charliecloud} offer alternative HPC containers but with tighter administrative coupling or limited features.
Among solutions that can be used to achieve automatic recovery, DMTCP~\cite{ansel2009dmtcp} and CRIU~\cite{andrijauskas2024criu} provide transparent process-level checkpointing although not compatible for capturing application-level agent reasoning state.
And for cluster management, Kubernetes~\cite{burns2016borg} is widely being used to manage long-running services, however it introduces unnecessary overhead and requires replacing existing HPC scheduler rather than leveraging it.

In summary, none of the existing approaches provide the required capabilities to seamlessly execute agentic workflows on existing HPC clusters out of the box.
RASER bridges this gap by leveraging Slurm's native primitives and established tools such as Apptainer.
Specifically, the combination of job arrays, the \texttt{requeue} mechanism, and container support proves sufficient for agent ensemble execution.
This approach requires no administrative coordination, persistent services, database infrastructure, or kernel privileges.

\section{RASER's Architecture and Implementation}

\begin{figure}[htbp]
\centering
\begin{tikzpicture}[
    box/.style={draw, rounded corners=3pt, minimum width=3.2cm,
                minimum height=1.4cm, align=center, text width=3cm,
                fill=#1, line join=round},
    arrow/.style={->, >=stealth, thick, line cap=round},
    label/.style={font=\bfseries\small},
    pillar/.style={font=\bfseries, rotate=90, anchor=south,
                   text width=2.5cm, align=center},
    innovation/.style={font=\small\bfseries, fill=orange!10,
                       rounded corners=2pt, inner sep=3pt,
                       draw=orange!50, line join=round},
    summary/.style={text width=3.2cm, align=center, font=\small}
]
\node[pillar] (p1) at (-5.5, 3.5) {Isolation};
\node[pillar] (p2) at (-6, 1.5) {Check-pointing};
\node[pillar] (p3) at (-5.5, -0.5) {Scheduling};

\node[label, color=red!60!black, anchor=west] at (-5.3, 5.2) {Existing approaches};

\node[box=red!8] (k8s) at (-3.5, 3.5) {%
    \textbf{Kubernetes / MPI}\\[2pt]
    orchestration\\[2pt]
    \textit{\small privileged spawn}};

\node[box=red!8] (extdb) at (-3.5, 1.5) {%
    \textbf{External database}\\[2pt]
    or CRIU daemon\\[2pt]
    \textit{\small state outside}};

\node[box=red!8] (static) at (-3.5, -0.5) {%
    \textbf{Static task-to-array}\\[2pt]
    mapping\\[2pt]
    \textit{\small rigid allocation}};

\draw[arrow, red!40!black] (k8s) -- (extdb);
\draw[arrow, red!40!black] (extdb) -- (static);

\node[summary, color=red!50!black] at (-3.5, -2) {%
    High privilege\\Coupled dependencies\\Fragile to failures};

\node[label, color=green!40!black, anchor=west] at (2.6, 5.2) {RASER};

\node[box=green!8] (app) at (3.5, 3.5) {%
    \textbf{Unprivileged}\\[2pt]
    Apptainer\\[2pt]
    \textit{\small bind-mounted runtime}};

\node[box=green!8] (self) at (3.5, 1.5) {%
    \textbf{Self-checkpointing}\\[2pt]
    agents\\[2pt]
    \textit{\small serialized + git}};

\node[box=green!8] (dyn) at (3.5, -0.5) {%
    \textbf{Work stealing}\\[2pt]
    queue\\[2pt]
    \textit{\small elastic within static}};

\draw[arrow, green!40!black] (app) -- (self);
\draw[arrow, green!40!black] (self) -- (dyn);

\node[summary, color=green!40!black] at (3.5, -2) {%
    User-level control\\Self-contained\\Fault resilient};

\draw[->, dashed, gray, thick] (-1.8, 3.5) -- (1.8, 3.5)
    node[midway, above=6pt, innovation] {Runtime injection};

\node[font=\small, text width=3cm, align=center] at (0, 2.7) {%
    No image modification required};

\draw[->, dashed, gray, thick] (-1.8, 1.5) -- (1.8, 1.5)
    node[midway, above=6pt, innovation] {App-level checkpoint};

\node[font=\small, text width=3cm, align=center] at (0, 0.7) {%
    No kernel privileges needed};

\draw[->, dashed, gray, thick] (-1.8, -0.5) -- (1.8, -0.5)
    node[midway, above=6pt, innovation] {Elastic load balance};

\node[font=\small, text width=3cm, align=center] at (0, -1.3) {%
    Within Slurm's static allocation};

\end{tikzpicture}
\caption{Existing approaches vs. RASER's for enabling agentic workloads on HPC}
\label{fig:architecture}
\end{figure}
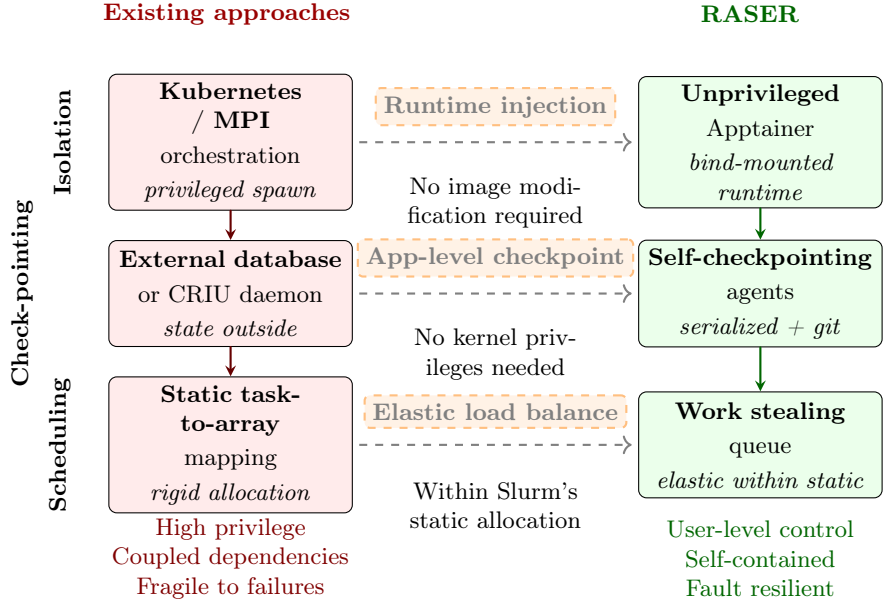

Various approaches attempt to leverage existing tools to enable agentic workloads on HPC clusters.
These approaches tackle the requirements of agentic workloads in three layers: Scheduling, Checkpointing, and Isolation.
However, using these approaches either requires privileged access to the HPC cluster, increases complexity and maintenance efforts by adding yet another layer of abstraction, or sacrifices resources and performance in favor of task automation.
RASER addresses these shortcomings and transforms Slurm from a static resource allocator into a resilient agentic platform.
Figure~\ref{fig:architecture} illustrates a visual comparison of RASER and existing approaches for enabling agentic workloads on HPC clusters.

RASER is designed to be simple, robust, and framework agnostic.
Its modular architecture ensures interoperability across different AI agent frameworks and various datasets.
While the main orchestrator has a generic design, framework-relevant agents and datasets can be supported as external plugins.
At initialization step, on the login node of HPC cluster, RASER parses the configurations, generates the shared task queue, loads required plugins according to users configuration, and submits a job array of agents.
Plugins bridge RASER and the target agent's runtime by handling the platform-specific setups, injecting agent-specific variables to the wrapper script, updating heartbeats, and if necessary patching the agent code.
For complex frameworks that do not expose native checkpointing hooks, plugins can utilize runtime injection (monkey-patching) to transparently wrap the agent's LLM query and tool execution loops with state-saving hooks.
This design enables RASER to execute workload-specific agent implementations, while keeping the main codebase generic.

Using RASER, user only needs to write a  plugin script (with \texttt{get\_tasks()} and \texttt{run\_task()}), while the entire complex distributed systems logic such as file locking (fcntl), state serialization to survive preemption (SIGTERM handling), and automated job recycling via \texttt{scontrol requeue} is handled natively and transparently by the orchestrator without manual intervention.

\begin{figure}
\includegraphics[width=\textwidth]{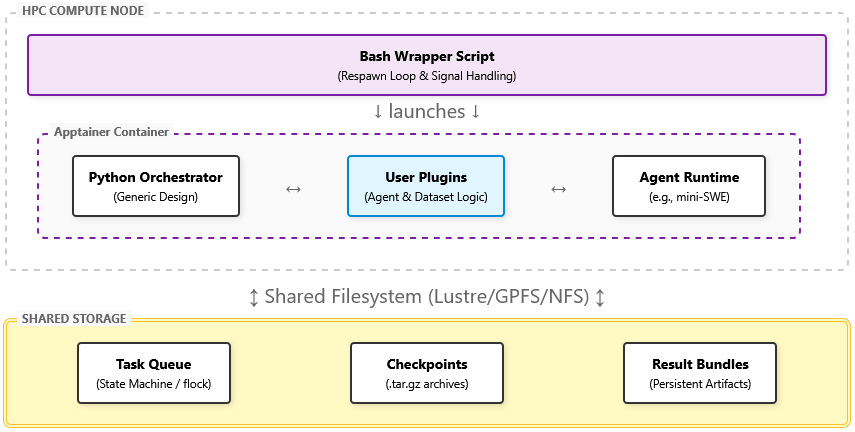}
\caption{Schematic architecture of RASER, illustrated at runtime on single node.}
\label{fig:raser-architecture}
\end{figure}

Figure~\ref{fig:raser-architecture} illustrates a schematic view of RASER architecture.
Each task of the submitted job array runs a persistent Bash wrapper script on a compute node.
This wrapper launches a container that hosts the Python orchestrator, passing in the agent’s workspace, queue path, and API keys via bind mounts and environment variables.
Thus, each agent effectively runs on a compute node, sand-boxed inside a container.
The wrapper itself resides in an infinite respawn loop.
If the Python process is killed for any reason, the container is destructed but the Bash script remains, logs the crash, briefly waits, and then relaunches the container; enabling an immortal long‑running agent loop.

RASER follows a zero image modification policy, that ensures user's containers require no RASER-specific software. 
To achieve this goal, an orchestration runtime is bind-mounted at execution time that allows code updates without image modifications.
Framework specific configurations and API keys are injected dynamically at runtime via \verb|--env| flags. 
In order to provide filesystem isolation which is particularly crucial for agents that modify shared resources such as git repositories, each agent operates on a private bind-mount folder\footnote{\texttt{\$RUN\_FOLDER/agents/\$SLURM\_ARRAY\_TASK\_ID/workspace}}, preventing cross-contamination, thus ensuring filesystem isolation at agent-level.
In addition, a shared filesystem\footnote{\texttt{/run\_data}} is mounted for maintaining atomic reading of task queues, storing agents logs, checkpoints and the results of the agents' work.
RASER also binds isolated \verb|--home| mount to each agent, redirecting its configuration and cache directories to a unique path such that multiple agents can run on the same node without filesystem collisions.

Unprivileged Apptainer~\cite{kurtzer2017singularity} containers are used within user namespace to provide full process isolation.
To survive container destruction during preemptions, RASER serializes each agent's state to the host filesystem as a persistent checkpoint.
By receiving a preemption signal (SIGTERM) due to node preemption or reaching the Slurm wall-time, the agent compresses its private workspace into a portable archive and stores it in the shared filesystem\footnote{\texttt{/run\_data/checkpoints/*.tar.gz}}, then exits with code 99; indicating a preemption state to the wrapper.
The wrapper will detect the preemption, thus calls the \texttt{scontrol requeue} to re-execute the preempted job and terminates itself.
Upon the agent's next execution, this checkpoint is restored and the agent resumes from its preemption point.
Once the queue is fully processed and the orchestrator finishes normally, the parent Python process exits with code 0 and the wrapper will break the loop, terminating the agent gracefully.

\begin{figure}
\includegraphics[width=\textwidth]{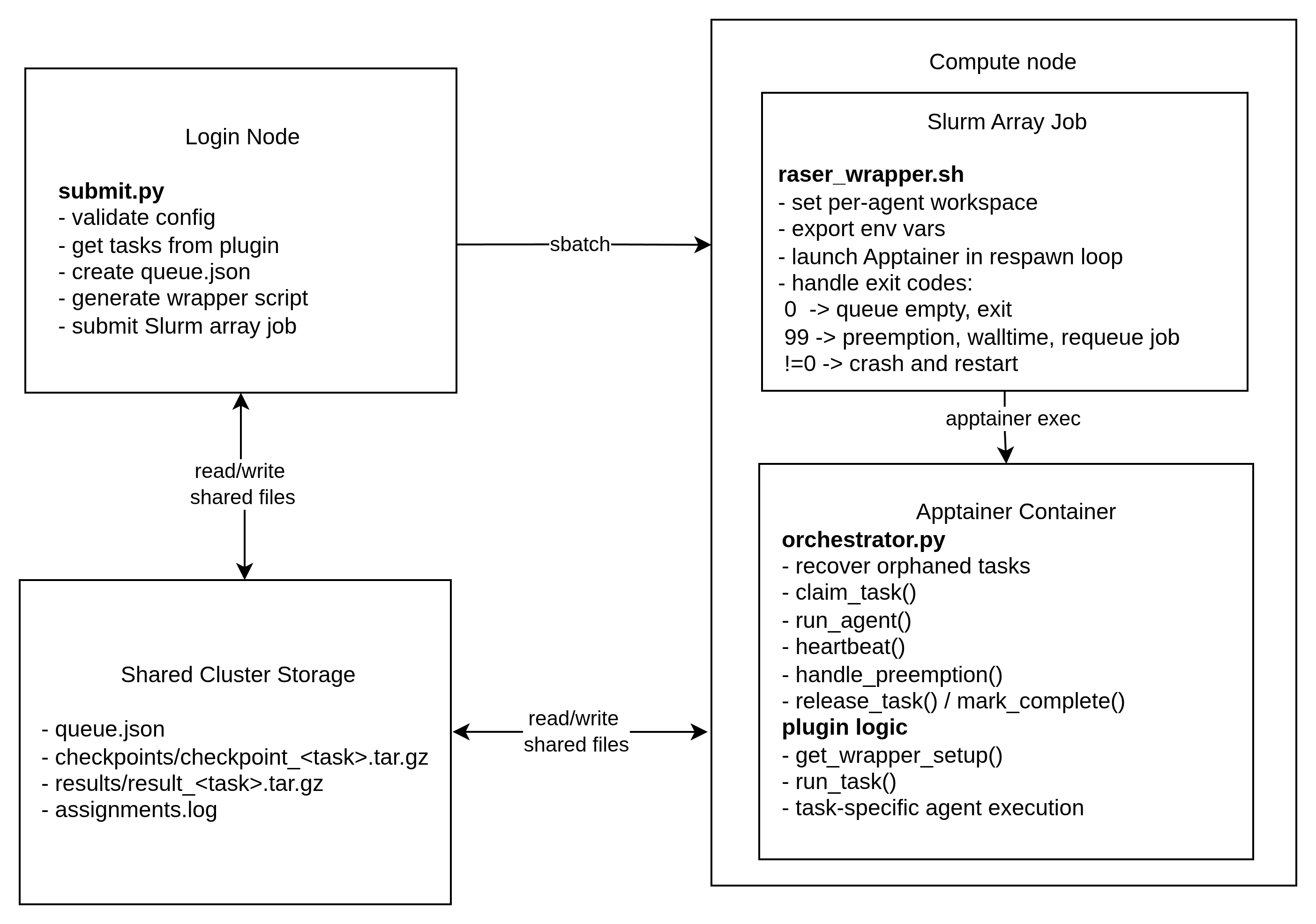}
\caption{RASER Workflow and Functions}
\label{fig:flow}
\end{figure}

In normal Slurm job arrays, workloads are statically partitioned and mapped to specific array indices prior to execution. 
RASER, in contrast, is designed to entirely decouple the Slurm array index from task assignment.
Load balancing is handled elastically at runtime via a centralized, JSON-based task queue residing on the shared filesystem, and a work stealing policy.
Agents claim tasks via atomic file locking using \texttt{fcntl}, polling the queue to claim \texttt{pending} tasks or reclaim \texttt{stale} tasks from unresponsive peers. 
The detailed control flow of RASER is shown in Fig~\ref{fig:flow}.
If an agent fails to update its timestamp for $T_{steal}$ seconds (by default 3 minutes), other agents may reclaim its task, enabling automatic load balancing without central coordination.
This uncoordinated work-stealing architecture ensures that faster or uninterrupted agents dynamically claim a larger share of the workload, effectively achieving elastic load balancing within the strict boundaries of Slurm allocation.

In order to further optimize resource utilization, a resource-aware, multi-tier task claiming approach is implemented.
Agents are provisioned across distinct resource tiers (in current implementation: Low-, Medium-, and High-memory) and filter the shared task queue based on each task's failure history.
When the orchestrator detects that an agent is crashed, it increments the value of a counter for that task before returning it to the \texttt{pending} pool.
Higher-tier agents prioritize tasks with elevated counters, while lower-tier agents are restricted to fresh tasks.
This dynamic escalation ensures that lightweight tasks do not monopolize expensive, high-memory agents, while memory-intensive workloads are systematically routed to more capable agents, thus maximizing overall efficiency.

A multi-layer fault tolerance mechanism improves RASER's resilience by triggering appropriate actions depending on whether a failure occurs at the task, agent, or job level.
Tasks may fail due to issues such as out-of-memory errors, infinite tool-calling loops, or third-party API failures.
Upon task failure, retries, task re-queuing, or reassignment to other agents will be triggered accordingly.
To maintain progress, agents running on preempted resources create checkpoints of their workspace and state before gracefully terminating; they are then automatically requeued and resume their task, ensuring a steady number of active agents as long as there are remaining tasks.
At the job level, most failures are handled transparently by Slurm through automatic job requeuing.
Although failures may prevent proper checkpointing of agents state and therefore, requiring agents to restart from scratch, RASER preserves the overall progress.
Thus, the only penalty is reexecution of the lost computation interval.
A summary of common causes of failure at each level and the triggered actions in each case is summarized in Table~\ref{tab:recovery}.
Current implementation of RASER assumes the following requirements on the underlying HPC cluster:
\begin{itemize}
    \item Apptainer~$1.1+$ with user namespace support (\verb|--userns|)
    \item POSIX filesystem with \verb|fcntl| support (e.g., Lustre, GPFS, NFSv4)
    \item Slurm~$20.02+$ with \verb|--requeue| and \verb|--signal| support
    \item Python~$3.8+$ (login node only)
\end{itemize}

\begin{table}
\caption{Failure recovery mechanisms in RASER.}
\label{tab:recovery}
\centering
\begin{tikzpicture}[
    every node/.style={font=\small, align=left},
    card/.style={rectangle, draw=gray!60, fill=white, 
                 inner sep=10pt, text width=\textwidth}
]

\path (0,0) node[card] (task) {
\textbf{Task level}\hfill\textit{$0$--$\mathcal{O}(T_{\text{steal}})$}\\
\textcolor{gray}{Common reason:} OOM, infinite loop, API error\hspace{2em}\\
\textcolor{gray}{Triggers:} Timeout-based reclamation
};

\path (task.south) ++(0,-0.6cm) node[card, anchor=north] (agent) {
\textbf{Agent level}\hfill\textit{$<60$\,s (restart + restore)}\\
\textcolor{gray}{Common reason:} SIGTERM (preemption), wall-time expiration\hspace{2em}\\
\textcolor{gray}{Triggers:} Immediate checkpointing $\to$ exit 99 $\to$ Slurm \texttt{requeue}
};

\path (agent.south) ++(0,-0.6cm) node[card, anchor=north] (job) {
\textbf{Job level}\hfill\textit{Job restart + restore}\\
\textcolor{gray}{Common reason:} Node failure\hspace{2em}\\
\textcolor{gray}{Triggers:} Slurm's automatic array \texttt{requeue}
};

\end{tikzpicture}
\end{table}

\section{Evaluation and Discussions}
RASER provides a powerful and accessible user-space framework that enables seamless agentic workflows on HPC clusters.
Through its resilience mechanism, it protects internal agent states and ensures that computational progress is maintained with minimal interruption.
The architecture's simplicity and integration with existing software stacks provides high robustness and low overhead.
Moreover, its multi-tier, non-coordinated load balancing mechanism optimizes resource utilization.
To the best of our knowledge, RASER is the only existing solution that provides a Slurm-native, installation-free, resilient agentic execution environment on HPC clusters, while performing distributed load balancing without requiring any privileged access.

Although a direct comparison by other frameworks is not possible, to demonstrate the effectiveness of RASER's approach, a subset of 60 tasks from SWE-bench-Lite~\cite{jimenez2024swebench} challenge was executed using the mini-SWE-agent~\cite{yang2024sweagent} framework.
Two distinct setups  were evaluated: (1)~static partitioning\footnote{Slurm job array with static task distribution.}, and (2)~RASER.
\begin{figure}
    \centering
    \includegraphics[width=0.88\linewidth]{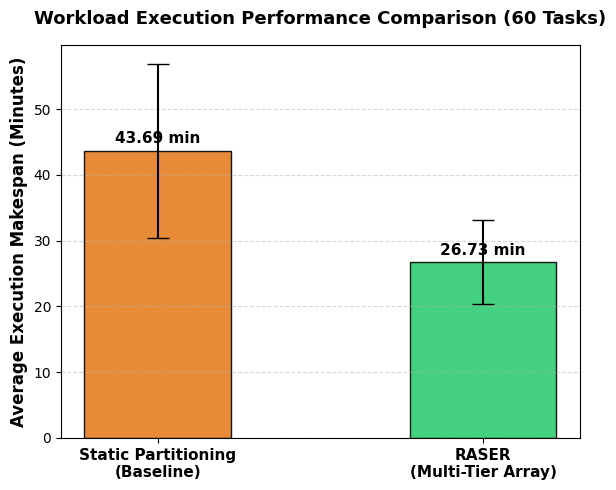}
    \caption{Comparing the average makespan between static partitioning and RASER}
    \label{fig:compare}
\end{figure}

In static partitioning scenario, Slurm job array with fixed task distribution was used to execute 15~concurrent agents.
At initialization, all 60~tasks were evenly assigned to the 15~agents (4~tasks per agent).
Once an agent completes its pre-assigned subset of tasks, it terminates and releases its allocated resources.
Similar to the previous setup, 2~CPUs and 2~GB of RAM were allocated to each agent.
We ran this test five times, and the average time that 15~agents managed to finish all 60~challenges was 43.69~minutes.
In the second setup, the same 60 tasks were executed using RASER. The framework initiated 15~concurrent agents with heterogeneous memory allocations: five agents with 1~GB, five agents with 2~GB, and five agents with 4~GB of RAM. We ran this evaluation five times; using the 3-tier queuing system and dynamic work stealing, the average makespan across all challenges was reduced to 26.73~minutes.
This represents a 38.8\%~reduction compared to static partitioning scenario.

It is worth mentioning that to verify the robustness of RASER in case of failures, Slurm's preemption was simulated via \verb|scancel| \verb|--signal=SIGTERM| at random intervals (fault injection).
RASER agents were practically immortal unless all tasks are fully processed and executed to completion.
Furthermore, despite the injected failures, the agents internal states were successfully restored after each failure.
However, since the amount of lost computation intervals caused by failures are highly task-dependent, these values are not reported in this section.

Based on the evaluations performed on Romeo\footnote{\url{https://compendium.hpc.tu-dresden.de/jobs\_and\_resources/romeo/}} HPC cluster, as can be seen in the figure~\ref{fig:compare}, RASER reduces the makespan\footnote{The total elapsed time from the moment the first task of a job (or a set of jobs) begins execution until the very last task is completed.} of agentic workflows nearly 39\% on heterogeneous workloads compared to static partitioning, while providing near-full CPU utilization.
On NVMe-backed file-systems the checkpoint restore overhead is estimated to be less than 30 seconds.
The sandboxing of agents using Apptainer containers also proved effective.
This is important to note that despite the use of containers by RASER, they maintain direct hardware access therefore, the overhead of virtualization is minimal.
Similar to any other parallel execution framework, the number of parallel processes (agents) and the duration of each process (task length) will influence the overall performance of RASER.
Since RASER does not impose any abstraction layer on the existing HPC software stack, the peak performance of RASER is similar to the underlying HPC environment.
The only dependency that RASER introduces is the shared filesystem operations.
However, given the negligible execution time of these operation in compare to LLM API calls, this dependency does not create any bottleneck. While no filesystem bottlenecks were observed during our current evaluation, more extensive scalability testing with a larger number of concurrent agents is required in future work to thoroughly guarantee performance under extreme filesystem lock contention. 

\section{Conclusion and Future Work}

This work presented RASER, a user-space framework that enables seamless execution of agentic workflows on production HPC clusters by extending Slurm's native primitives.
RASER combines job arrays with work stealing via shared filesystem queues, application-level checkpointing for fault tolerance, and container-based isolation.
The proposed solution enables large-scale AI agent evaluation and deployment on existing HPC clusters, bridging the gap between cloud-native orchestration and traditional batch scheduling.
Evaluations demonstrate significant reduction in makespan compared to static partitioning while achieving near-full CPU utilization.
RASER's installation-free, user-space design requires no kernel privileges or external database infrastructure, making it an accessible solution for deploying agentic workflows on existing HPC clusters.
For future work, two main areas require further investigation.
First, implementing task-relevant model switching capabilities remains challenging due to two fundamental issues: smaller models may not comprehend the complex reasoning state constructed by larger models, leading to hallucination, and different models employ incompatible formatting and tool-call conventions.
Second, a fundamental challenge in distributed systems which is also present here is distinguishing between an LLM that is genuinely reasoning and one that has entered a hung state.
Proper identification of these states further improves the effectiveness of job stealing strategies.
Finally, while our current evaluation relies on CPU-only allocations making external API calls, RASER’s workload-agnostic design can theoretically support agents requiring hardware acceleration. Integrating and benchmarking GPU provisioning within our multi-tier architecture also remains an area for future investigation.

\begin{credits}


\subsubsection{\discintname}
The author(s) used tools powered by MinMax-M2.5, to improve spelling, grammar, clarity, and readability during manuscript preparation.
\end{credits}

%
%
%
\bibliographystyle{splncs04}
\bibliography{references}
\end{document}